\documentclass[twocolumn,superscriptaddress,aps,prb,showpacs]{revtex4}
\usepackage{graphicx,amsmath,amssymb}
\usepackage{epstopdf}
\usepackage{amssymb}
\usepackage{grffile}
\usepackage[usenames]{color}
\usepackage{indentfirst}
\usepackage{float}
\usepackage{color}
\usepackage{mathrsfs}
\usepackage{subfigure}
\usepackage{dcolumn}
\usepackage{bm}
\usepackage[colorlinks=true,bookmarks=false,citecolor=blue,linkcolor=red,urlcolor=blue]{hyperref}

\begin{document}
\title{The entanglement entropy of the quantum Hall edge and its geometric contribution}
\author{Dan Ye$^{a}$}
\affiliation{Department of Physics and Chongqing Key Laboratory for Strongly Coupled Physics, Chongqing University, Chongqing 401331, People's Republic of China}
\author{Yi Yang$^{a}$}
\thanks{$^{a}$These authors have contributed equally to this work.}
\affiliation{Department of Physics and Chongqing Key Laboratory for Strongly Coupled Physics, Chongqing University, Chongqing 401331, People's Republic of China}
\author{Qi Li}
\affiliation{GBA Branch of  Aerospace Information Research Institute, Chinese Academy of Sciences, Guangzhou 510535, People's Republic of China}
\author{Zi-Xiang Hu}
\email{zxhu@cqu.edu.cn}
\affiliation{Department of Physics and Chongqing Key Laboratory for Strongly Coupled Physics, Chongqing University, Chongqing 401331, People's Republic of China}

\begin{abstract}
Generally speaking, the entanglement entropy (EE) between two subregions of a gapped quantum many-body state is proportional to the area/length of their interface due to the short range quantum correlation. However, the so-called area law is violated logarithmically in a quantum critical phase. Moreover, the subleading correction exists in a long range entangled topological phases. It is referred to as topological EE which is related to the quantum dimension of the collective excitation in the bulk. Further more, if a non-smooth sharp angle is in the presence of the subsystem boundary, a universal angle dependent geometric contribution is expected to appear in the subleading correction. In this work, we simultaneously explore the geometric and edge contribution in the integer quantum Hall (IQH) state and its edge reconstruction in a unified bipartite method. Their scaling is found to be consistent with the conformal field theory (CFT) predictions and recent results of the particle number fluctuation calculations.
\end{abstract}
\date{\today }

\maketitle

\section{Introduction}
Quantum entanglement is a fundamental and important tool to probe the properties of a variety of  physical systems such as black hole in the astrophysics, quantum phase transition in condensed matter physics and photosynthesis in biophysics ~\cite{RMPVedral}.  In a bipartite system, one  usually calculate the von Neumann entropy or the $\alpha$-R\'enyi entropy to quantitatively describe the magnitude of the entanglement between two subsystems. Once the system size is smaller than the correlation length, the entropy is proportional to the volume of system. For a gapped state, it is generally proportional to the area/length of the interface between two subsystems. This is referred to as the area law in three-dimensional or perimeter law in two-dimensional system. Heuristically, this could be understood from the facts that an energy gap gives rise to a finite correlation length which defines the scale on which particles inside the subsystem are correlated with the environment. In gapless critical systems, such as the quantum Hall edges or critical spin systems which could be described by the conformal field theory (CFT), it is known that the EE has a logarithmic dependence on the boundary length and the prefactor is related to the central charge of its underlying CFT~\cite{Holzhey,Vidal,Calabrese,WenIJPM}.

After the discovery of topological quantum systems, such as the fractional quantum Hall effects~\cite{Tsui,Laughlin83}, it is known that there is an extra correction of the bulk EE which depends on the quantum dimension of the collective excitation in the bulk. It is referred to as the topological EE $\gamma$~\cite{Kitaev,Levin}, as an important quantity to characterize the nontrivial topology of the long range entangled quantum many-body states. Moreover, Li and Haldane~\cite{HaldaneES} found that the eigenvalue spectra of the reduced density matrix, named the entanglement spectrum provides more information about the topology since it could be treated as a virtual edge excitation spectra at the bipartite boundary. The mechanism of the bulk-edge correspondence ~\cite{Chandran,Luo} could tell us many of the the bulk properties. On the other hand, the quantum Hall edge excitation are usually chiral gapless mode which could be described by $(1+1)$d chiral CFT. Once the cutting line is along the realistic quantum Hall edge with length $l_A$, a logarithmic type of $\alpha$-R\'enyi EE $S_{\text{edge}} \simeq \frac{c+\bar{c}}{12} (1+\frac{1}{\alpha})\log l_A$  ~\cite{Estienne20} with central charge $c$ is expected. For the chiral edge mode, the anti-holomorphic part $\bar{c} = 0$ and thus $S_{\text{edge}} \simeq \frac{c}{12}(1+\frac{1}{\alpha})\log l_A$.

Up to now, the behavior of the EE mentioned above are under assumption that the bi-partition has a smooth boundary, such as a circle or an infinite straight line. Once the boundary having a sharp corner, regardless of whether the system is gapped or not, it was found that the corner on the boundary has an important contribution in the EE which was previously explored~\cite{r1,r2,r3,r4,r5,r6,r7,r8,r9,r10,r11,r12,r13,r14} in the two dimensional quantum critical systems and CFT. Recently, it was extended to the gapped topological system such as the integer quantum Hall states~\cite{Estienne20,Rozon,Sirois21}. The corner angle dependence of the EE is found to be universal~\cite{Estienne22}. Therefore, the complete formula of the EE is
\begin{eqnarray} \label{AllS}
S = a A + b\partial A +\gamma + S_{edge} + S(\theta) + \mathcal{O}(1/L).
\end{eqnarray}
in which we include the volume, area, topological, edge and corner contributions in the first five terms.

In this work, as an example of unification, we consider the quantum Hall state in disk geometry with an open boundary. The fan-shaped bipartite EE with different radius  simultaneously gives the contributions from the area law, critical edge mode and the non-smooth corner. For the integer quantum Hall state, we observed that the corner contribution has the similar behavior of the charge cumulation at the cone tip if we put the electrons in a cone shaped geometry. Moreover, the logarithmic behavior and the central charge are obtained and the results are immune from the edge reconstruction which conserves the chirality.  This could be an explanation of recent observation that a universal the angle dependence of the particle number fluctuations in a non-smooth bi-partition.

The rest of the paper is organized as follows. In section II, we revisit the correlation matrix and EE with real space cut in disk geometry. The exact prefactor of the area law is found for $\alpha-$R\'enyi entropy with $\alpha = 1,2,3$. In section III, the corner contribution is obtained after subtracting the are law part by a fan-shaped cut in the bulk. A cone shaped quantum Hall state reveals the charge cumulation at the cone tip which has the similarity of the EE. In Sec IV, we consider the fan-shaped bi-partition including the quantum Hall edge. The logarithmic edge contribution could be obtained after subtracting both the area law and coner contribution. The central charge is found robust to any edge reconstructed pattern. Sec. V gives the conclusions and discussions.

\section{The entanglement entropy and area law}
For a two-dimensional electron gas in a strong perpendicular magnetic field, the typical length scale is the magnetic length $\textit{l}_B = \sqrt{\hbar/eB}$ which we set to one in the following. The single electron wave function in symmetric gauge are
\begin{eqnarray}
	\phi_{n,m}&=&c_{n,m}^\dagger|0\rangle\nonumber \\
	&=&(-1)^n\sqrt{\frac{n!}{2\pi2^{m}(m+n)!}}L_n^m(\frac{|\mathbf{z}|^2}{2})\mathbf{z}^{m}e^{-|\mathbf{z}|^2/4}
\end{eqnarray}
where $n,m$ are the Landau level and the angular momentum index respectively. In the lowest Landau level with $n=0$, the $m$th orbit is a Gaussian wave package in radial direction which has the most probable radius at $r_m = \sqrt{2m}$. For a bipartite system, the von Neumann entropy is defined as $S_1(\rho_A) = -\text{Tr}\rho_A\ln\rho_A$ once we have the reduced density matrix $\rho_A$ for the subsystem. More generally, the $\alpha$-R\'enyi entropies~\cite{Sirois21} are defined as $S_\alpha(\rho_A) = \frac{1}{1-\alpha} \ln(\text{Tr}\rho_A^\alpha)$ which reduces to the von Neumann entropy in the limit $\alpha \rightarrow 1$. Their scaling behaviors with increasing the size of the subsystem $A$ has yielded plentiful interesting results for both gapped and critical systems. Generally, the universality of the entanglement appears when the system length scale, such as the boundary length is much larger than $l_B$. Therefore, one usually suffers a strong finite size effect for small system sizes.  For a many-body system, diagonalizing the $\rho_A$ is usually limited to small systems because of the exponential explosion of its dimension as increasing the system size. Fortunately, for a non-interacting fermionic system which has Slater determinants as its eigenstates, it is known~\cite{Vidal,Peschel} that this could be simplified to calculate the eigenvalues of single particle correlation matrix $C_{ij} = \text{Tr}(\rho_A c_i^\dagger c_j)$ where $c_i$ is the single particle operator. Its dimension is the number of orbitals which linearly grows as increasing the system size. Furthermore, the correlation matrix is naturally diagonal in case the bipartite cutting conserves the symmetry of its parent wave function, i.e., the circular cutting on a disc or latitude cut on a sphere. In this case, the two types of entropy are defined by its diagonal terms, or its eigenvalues $\{\lambda_m\}$ as following
\begin{eqnarray}
S_1 &=& \sum_m [-\lambda_m \log \lambda_m -(1-\lambda_m)\log(1-\lambda_m)] \\
S_\alpha &=& \frac{1}{1-\alpha} \sum_m \log[\lambda_m^\alpha + (1-\lambda_m)^\alpha].
\end{eqnarray}

\begin{figure}[ht]
\center{\includegraphics[width=8cm]  {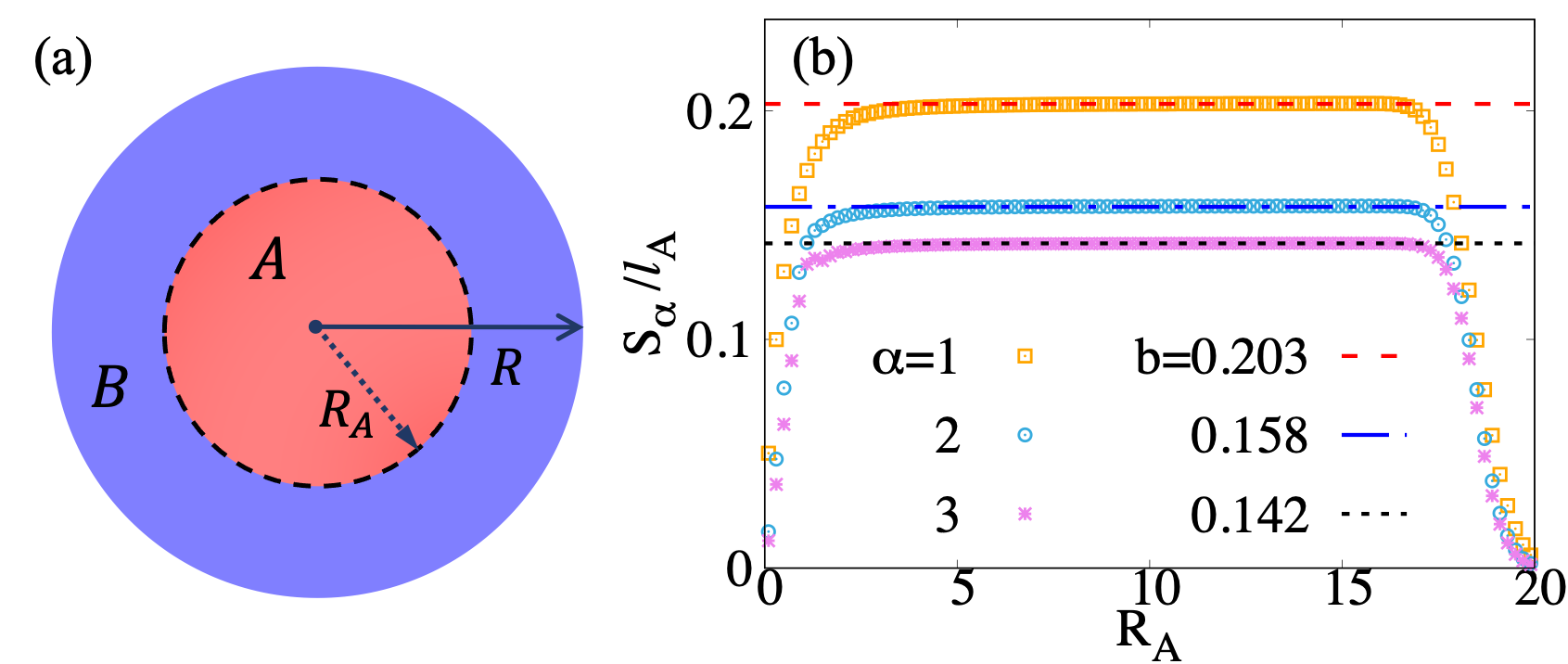}}
\caption{\label{fig1} (a) The smooth circular cut with radius $R_A$ in a finite disk. (b) The prefactor of the area law for $\alpha$-R\'enyi entropies. The well developed plateau appears when $R_A$ is larger than $1.0 l_B$ and smaller than the radius of system.}
\end{figure}

For a circular bipartite finite disk as shown in Fig.~\ref{fig1}(a), the electron operator for the $m$th orbital in the lowest Landau level ($n=0$) could be written as~\cite{Sterdyniak12,Dubail12,Simon12}
\begin{equation}
 c_m =  \alpha_{m}A_{m} + \beta_{m}B_{m}
\end{equation}
where $A_m$ and $B_m$ are the electron operators in subsystem A and its environment B. $\alpha_{m}^2$ and $\beta_{m}^2$ are the probabilities in two subsystems. For circular cut with radius $R_A$, they are
\begin{eqnarray}
\alpha^{2}_m = \int^{R_A}_0\int^{2\pi}_0|\phi_{0,m}(r,\theta)|^2d^2r = 1-\frac{\Gamma(1+m,\frac{R_A^2}{2})}{\Gamma(1+m)}
\end{eqnarray}
 and $\beta^{2}_{m} = 1 - \alpha^{2}_{m}$. For the $\nu=1$ IQH state  which does not have topological term $\gamma$, if $R_A$ is much smaller than that of the whole disk, namely the cut edge is far away from the physical edge at $R = \sqrt{2N_e}$, the EE contains only the area law term in Eq. (\ref{AllS}) as $S_{\alpha} = b_{\alpha} l_A + \mathcal{O}(1/l_A)$ where $l_A = 2 \pi R_A$ is the perimeter of the boundary. In Fig.~\ref{fig1}(b), we plot the $b_{\alpha} = S_{\alpha}/l_A$ as increasing the $R_A$ for system with $N_e = 171$ electrons. The prefactor $b_\alpha$, shown as constant values in a large range for $R_A > 1$ and $R_A < \sqrt{2N_{orb}}\simeq 18.5$, demonstrates a perfect linear behavior of $S_{\alpha}$ for a smooth cut in the bulk. The prefactor for $S_1$ is $b_1 \simeq 0.203$ which is exactly the same as that from a similar study in cylinder geometry where an analytical formula was given as $b_1 = \int \frac{d\mu}{2\pi} H[\frac{1}{2}\text{Erfc}(\mu)]\simeq 0.20329081$ in which the integral function is $H(x) = -x \log(x) - (1-x) \log(1-x)$~\cite{RodPRB}.  We found that the prefactors of the  $S_2$ and $S_3$ are $b_2 = 0.158$ and $b_3 = 0.142$ respectively. Their analytical results could also be obtained by the corresponding integrate function $H_{\alpha}(x) = \frac{1}{1-\alpha}\log(x^{\alpha} + (1-x)^{\alpha})$ which gives $b_2 = 0.15843$ and $b_3 = 0.14213$. Moreover, the $S_{\alpha =2,3}/l_A$ saturates faster than $S_1/l_A$ at small $R_A$ means the $\alpha$-R\'enyi entropy suffers weaker finite size effects than the Von Neumann entropy. As a conclusion in this section, we demonstrate that the area law prefactor $b_{\alpha}$ for IQH are the same for different geometries and could be calculated analytically.  This will be applied in the following to subtracting the area law term in the non-smooth cut.

\section{The corner contribution}

 \begin{figure}[ht]
\center{\includegraphics[width=8cm]  {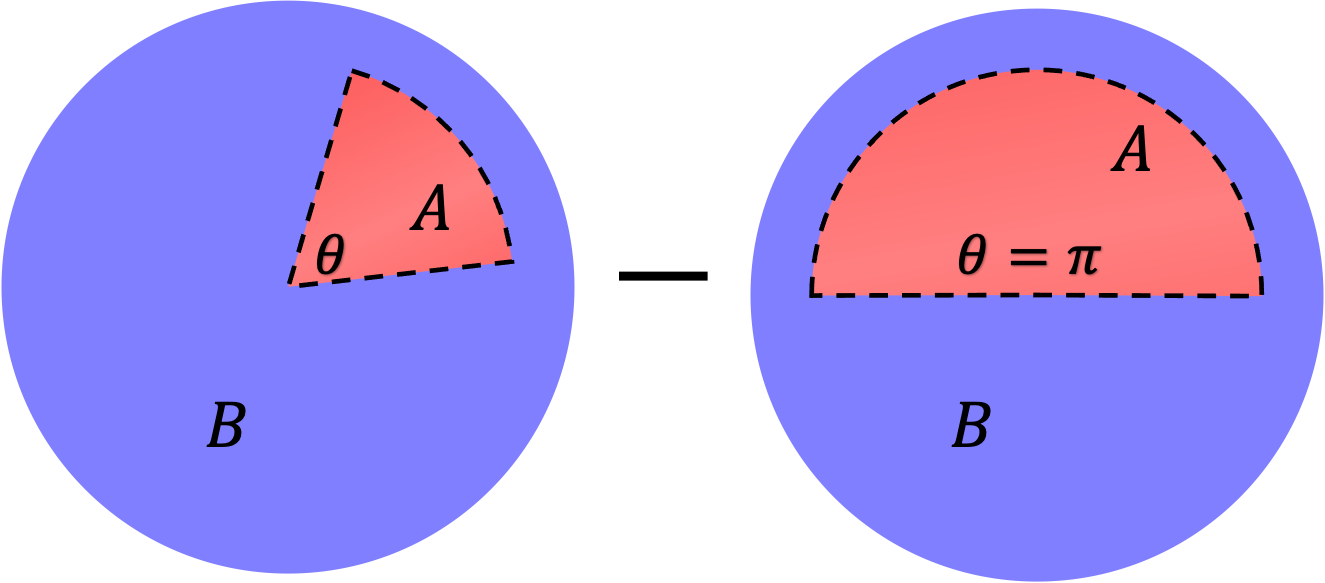}}
\caption{\label{fig2} The non-smooth cutting schematic diagram of a bipartite finite disk and the subtraction rule is applied to extract pure angle contribution.}
\end{figure}

 In this section, we consider a non-smooth partition as shown in Fig.~\ref{fig2}. The subsystem A is a fan-shaped region with a corner angle $\theta$ at the center of the disk. Supposing the arc-shaped boundary is far away from the physical edge of the disk, then the EE still in the area law region. The non-smooth cutting gives one corner of $\theta$-degree at the center and two corners of $\frac{\pi}{2}$-angle at the intersections with arc. Therefore, the EE in this case is
 \begin{eqnarray}
  S_\alpha = b_\alpha l_A + S_\alpha(\theta) + 2S_\alpha(\frac{\pi}{2}) + \mathcal{O}(1/l_A)
 \end{eqnarray}
 with the boundary length $l_A = 2 R_A + \theta R_A$. In order to screen out the pure angle contribution $S_\alpha(\theta)$, we subtract the value at $\theta = \pi$ with the same $R_A$. In this case, we have $S_\alpha(\pi) = 0$ and $l_A = 2 R_A + \pi R_A$. Therefore the residual entropy is $\Delta S_\alpha = b_\alpha (\theta - \pi) R_A + S_\alpha(\theta)$ in which the first term could be exactly obtained from the previous section supposing the area law and corner contributions are independent to each other.

 \begin{figure}[ht]
\center{\includegraphics[width=8cm]  {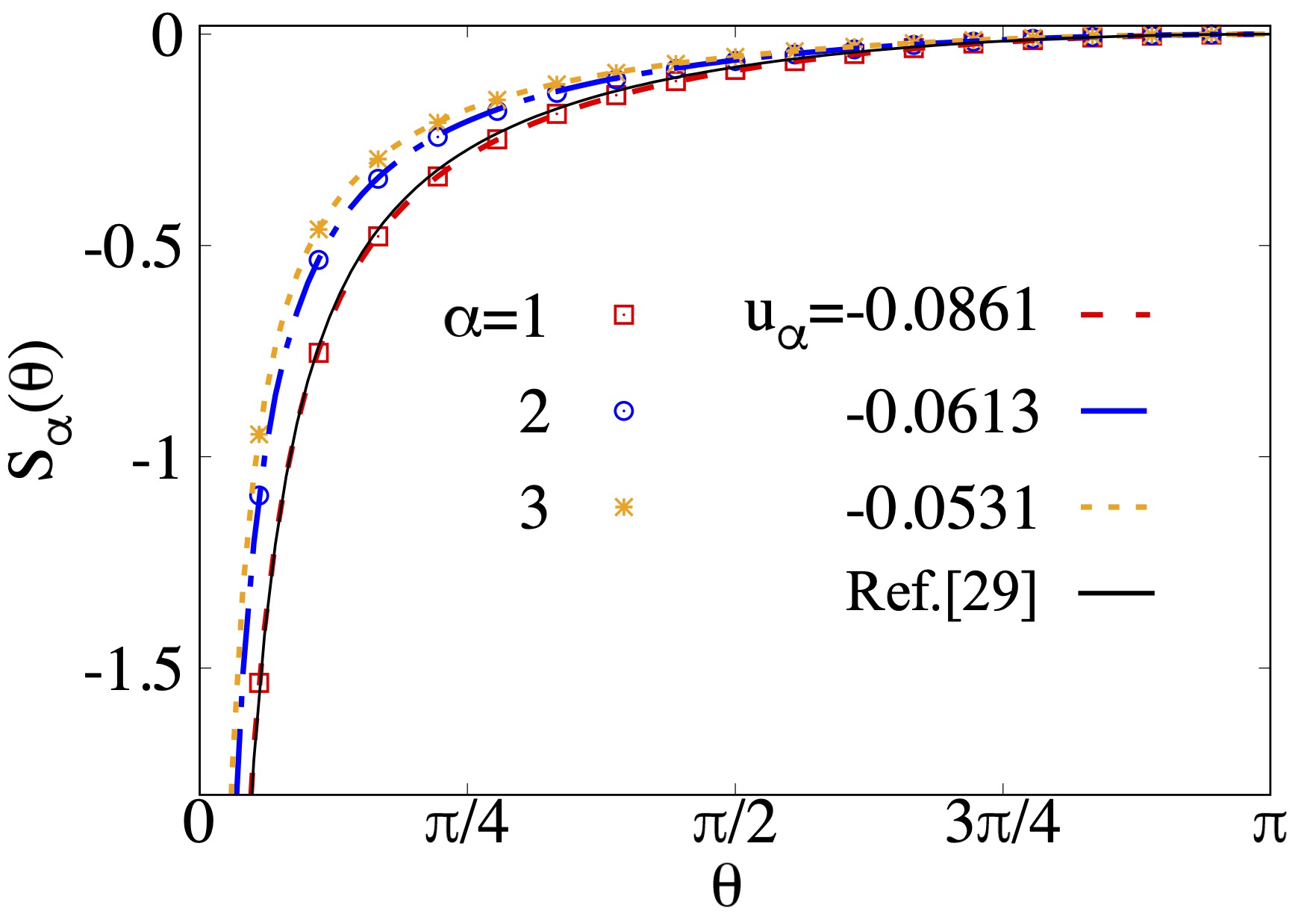}}
\caption{\label{fig3}  The $\theta$-dependence of the corner contribution in R\'enyi entropy for $171$ electrons.  After fitting the data by $S_{\alpha}(\theta)=u_{\alpha}[1+(\pi-\theta)\cot(\theta)]$, we find that $u_1=-0.0861(\pm 0.0009), u_2=-0.0613(\pm 0.0007)$ and $u_3=-0.0531(\pm 0.0006)$ which are consistent with the conjecture $S_\alpha(\theta) = (1+\frac{1}{\alpha}) f(\theta)$.}

\end{figure}

 In Fig.~\ref{fig3}, we plot the $S_\alpha(\theta) = \Delta S_\alpha - b_\alpha (\theta - \pi) R_A$ as a function of the angle $\theta$. A recent work of Ref.~\onlinecite{Estienne20} discussed the similar  bipartition by calculating the R\'enyi entropy via the cumulants of the particle number in subsystem $S_{\alpha} = \sum_m s_{\alpha}(m) \kappa_{2m}$ where $\kappa_m$  are the even cumulants of the particle number distribution in region A. In particular, $\kappa = \kappa_2$ is the variance of the number of particles in region A which was obtained analytically for IQH case, namely $\kappa(\theta) = \frac{\sqrt{N_e}}{\pi^{3/2}} + \frac{1}{2\pi^2}\ln(\sqrt{N_e}\sin(\frac{\theta}{2})) - \frac{1+(\pi-\theta)\cot\theta}{4\pi^2}$. The third term is the corner contribution in the second cumulant.   We assume the final $S_\alpha$ obeys the same $\theta$-dependence although the prefactor could be non-analytical. We fit the $S_\alpha(\theta)$ ~\cite{Hung14,Bueno15} with the function  $S_{\alpha}(\theta)=u_{\alpha}[1+(\pi-\theta)\cot(\theta)]$ and find that $u_1= -0.0861(\pm 0.0009)$, $u_2 = -0.0613(\pm 0.0007)$ and $u_3 = -0.0531(\pm 0.0006)$. In Ref.~\onlinecite{Rozon}, the corner contribution  was recently calculated in cylinder geometry. While $\theta \rightarrow 0$, it was found that the divergence of the $S_1(\theta)$ behaves as  $S_1(\theta) = -0.0886(\pm 0.0004)/\theta$ where the coefficient is consistent to our result of $u_1$ for the bulk corner. \textcolor[rgb]{1,0,0}{ It is interesting to know that a refined fit formula was also proposed~\cite{Sirois21}  which gives accuracy fits at both asymptotic limits, namely $S_\alpha(\theta) \simeq \beta_1 \frac{(\pi-\theta)^2}{\theta(2\pi-\theta)}-\beta_2[1+(\pi-\theta)\cot\theta]$ which has an extra $\theta$-dependent term. In our fitting process, the asymptotic behavior in the limit of $\theta \rightarrow 0$ gives $S_{\alpha} \simeq u_\alpha \pi / \theta$.  The coefficient $u_\alpha \pi \simeq 0.2705$ which is qualitatively consistent to the result of $0.276$ in Ref.~\onlinecite{Sirois21}.  The accuracy in this work could be lower due to the missing of the correction term and possible mixing different types of EE on a finite disk. }

 \begin{figure} [ht]
\center{\includegraphics[width=8cm]  {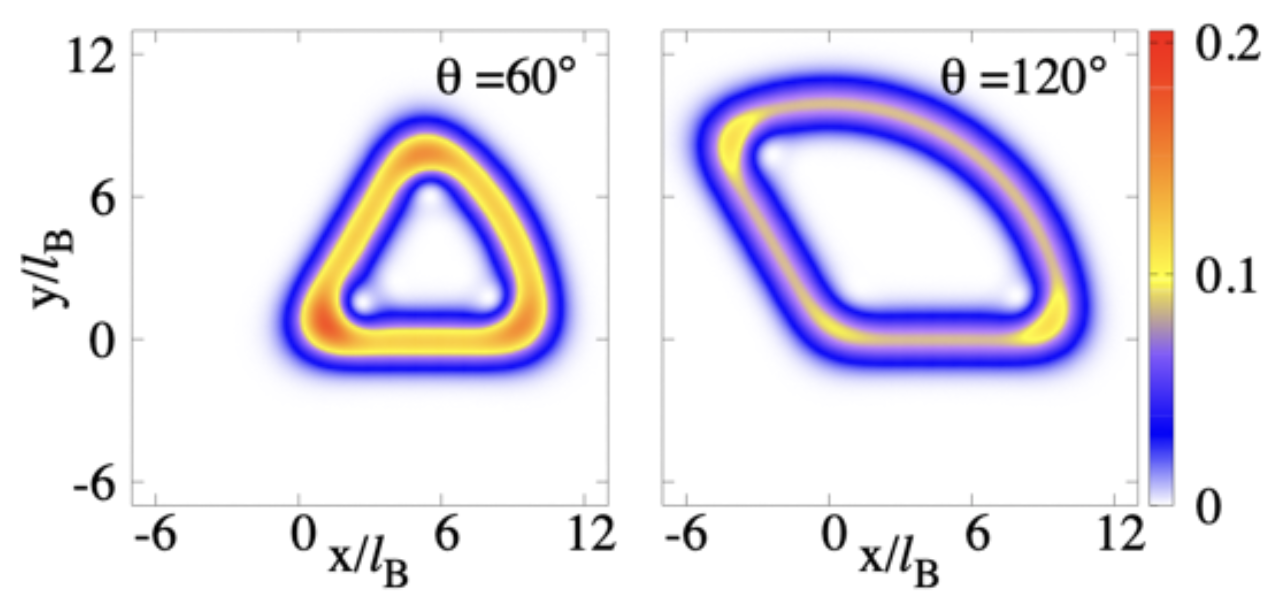}}
\caption{\label{fig4}  The single particle density for the state which has eigenvalue with the smallest $|\lambda - \frac{1}{2}|$. We choose $R_A=10l_B$, which is in bulk of a finite disk (with the physical edge $R=\sqrt{342}l_B$). We compare two cases with $\theta = \pi/3$ and $\theta = 2\pi/3$ and use the same color bar in two cases.  }
\end{figure}

 To look into more clearly the corner contribution of the EE, we treat the correlation matrix as entanglement Hamiltonian. The eigenstate which has the most important contribution in the EE is the one that has eigenvalue near $1/2$.  In Fig.~\ref{fig4}, we plot the single particle density for the state which has eigenvalue with the smallest $|\lambda - \frac{1}{2}|$. It is interesting to see that for this state, the
density mainly concentrates near the boundary of the subsystem A and the sharp corner has a higher density than the smooth edge. We compare two cases with $\theta = \pi/3$ and $\theta = 2\pi/3$. It is obvious that the acute angle corner has a much higher density than that of the obtuse angle corner. The behavior of the EE density accumulation at the sharp corner is qualitatively the same as the phenomena of the tip charge accumulation in electromagnetism.

\begin{figure}[ht]
\center{\includegraphics[width=8cm]  {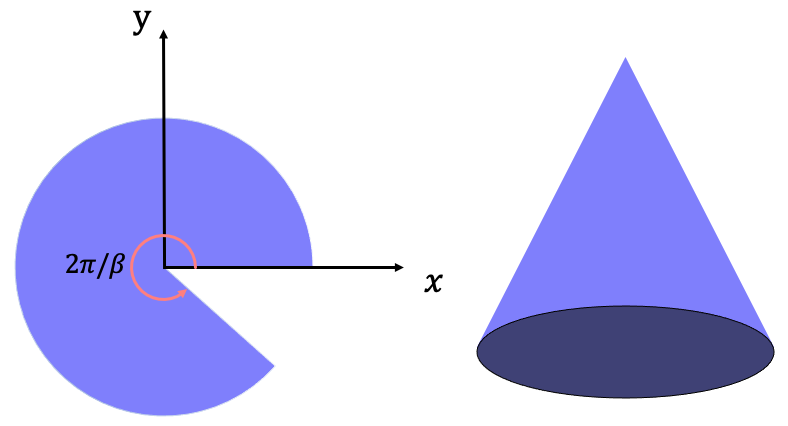}}
\caption{\label{fig5} A cone can be obtained from a fan-shaped geometry after gluing two edges together.}
\end{figure}

 Now we consider a realistic system with a corner. We suppose that the particles live in a fan-shaped geometry and glue it to a cone as shown in Fig.~\ref{fig5}. The quantum Hall state on a  cone has been realized experimentally in synthetic Landau levels for photons~\cite{SchineNature,Schine19}.  In this case, the Landau wave functions~\cite{Bueno12,Biswas16,Can16,Wu17} become to be
 \begin{eqnarray}
	\phi_{n,m}(\beta, \mathbf{z}) =\mathcal{N}_{n,m} L_n^{\beta m}(\frac{|\mathbf{z}|^2}{2})\mathbf{z}^{\beta m}e^{-|\mathbf{z}|^2/4}\nonumber
\end{eqnarray}
where $\mathcal{N}_{n,m}=(-1)^n\sqrt{\frac{\beta n!}{2\pi2^{\beta m}\Gamma(\beta m+n+1)}}$. The angle of the system $\theta = 2\pi/\beta$ could be continuously tuned by varying the parameter $\beta$. In this geometry, the wave function of the IQH state is
\begin{eqnarray}
|\Psi\rangle = \prod_{i<j}(z_i^{\beta}-z_j^{\beta}) \exp(-\sum_i |z_i|^2/4)
\end{eqnarray}

\begin{figure}[ht]
\center{\includegraphics[width=8cm]  {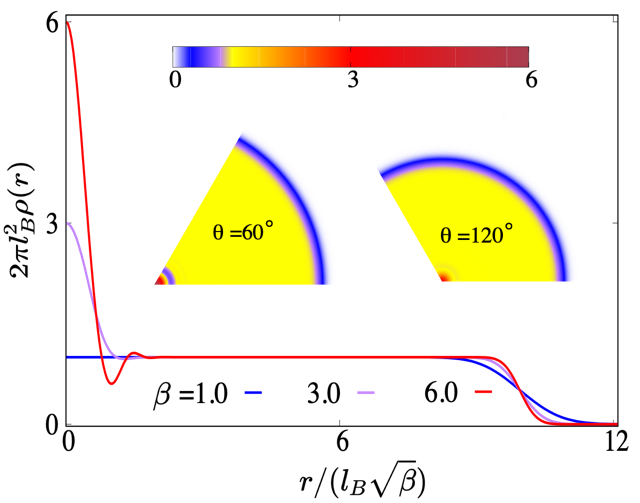}}
\caption{\label{fig6}  The radial density of the IQH state for $50$ electrons with $\beta$s.  $\beta = 1$ corresponds to the original disk geometry which has density $\rho(r) = \frac{1}{2\pi l_B^2 }$ in the bulk.  The two inserted plots are the 2D density profiles for different $\theta$s.}
\end{figure}
\begin{figure}[ht]
\center{\includegraphics[width=8cm]  {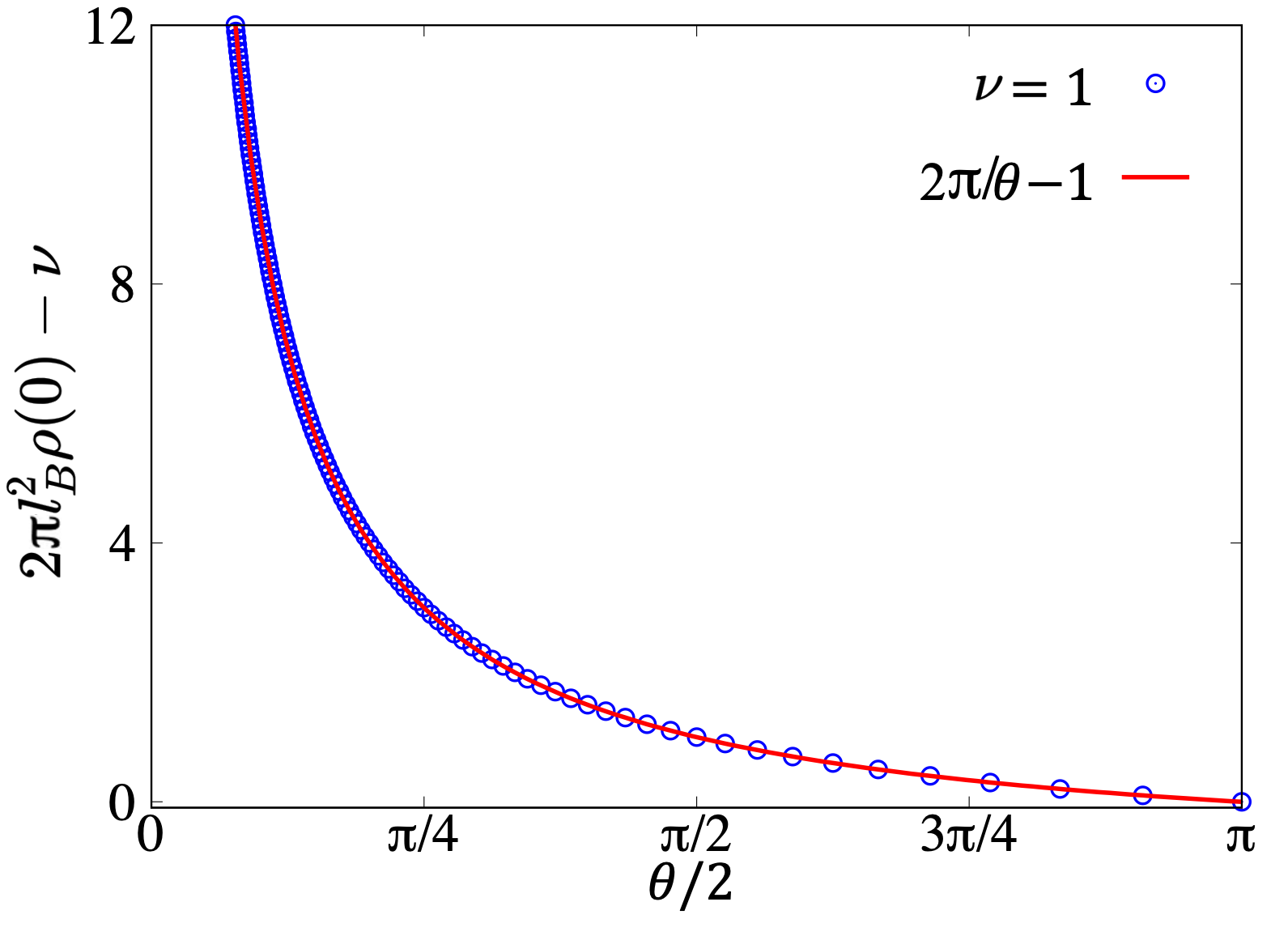}}
\caption{\label{fig7} The accumulation of density at the tip for IQH state $\rho(0)$. The data is perfectly fitted by $f(\theta) = 2\pi/\theta - 1$.}
\end{figure}

In Fig.~\ref{fig6}, we plot the radial density of the IQH state for different $\beta$s. $\beta = 1$ corresponds to the original disk geometry which has density $\rho(r) = \frac{1}{2\pi l_B^2 }$ at the center. While increasing $\beta$, or decreasing the $\theta$-angle of the system, the density at the corner tip increases dramatically. From the inserted two-dimensional density profiles, it is obvious that the density at the corner tip cumulates gradually and finally separates from the bulk. Since the occupation number for each orbital is exactly $n_m = 1$ for $\nu=1$ IQH. It is easily to analytically calculate the density at the center $\rho(0) =  \sum_m n_m |\phi_{0,m}(\beta, 0)|^2$ which is shown in Fig.~\ref{fig7}.  After subtracting the background density with $\nu = 1$, the data is perfectly fitted by $f(\theta) = 2\pi/\theta - 1$ which is the same as that of the EE while $\theta \rightarrow 0$.  Moreover, because of the Pauli exclusive principle of the fermions, the occupation number on each orbital is at most equal to  one.  Therefore, for other quantum Hall states, such as the fractional quantum Hall states which have $\nu < 1$ at $\beta = 1$, the $\theta \rightarrow 0$ behavior should be universal once the particle number on the $0$th orbital reach to one.

Therefore, we obtain the exact corner contribution of the EE via the fan-shaped bi-partition in the bulk. We found the $1/\theta$ divergence of the corner contribution near $\theta \rightarrow 0$ which has similarity to the charge density cumulation at the tip once we put the system on a cone.The similarity between the EE and the local charge density or its fluctuation has been studied in several systems, either classical or quantum many-body system~\cite{Estienne22,Jiang}.

 \section{The edge contribution and its universality}
 \begin{figure}[ht]
\center{\includegraphics[width=8cm]  {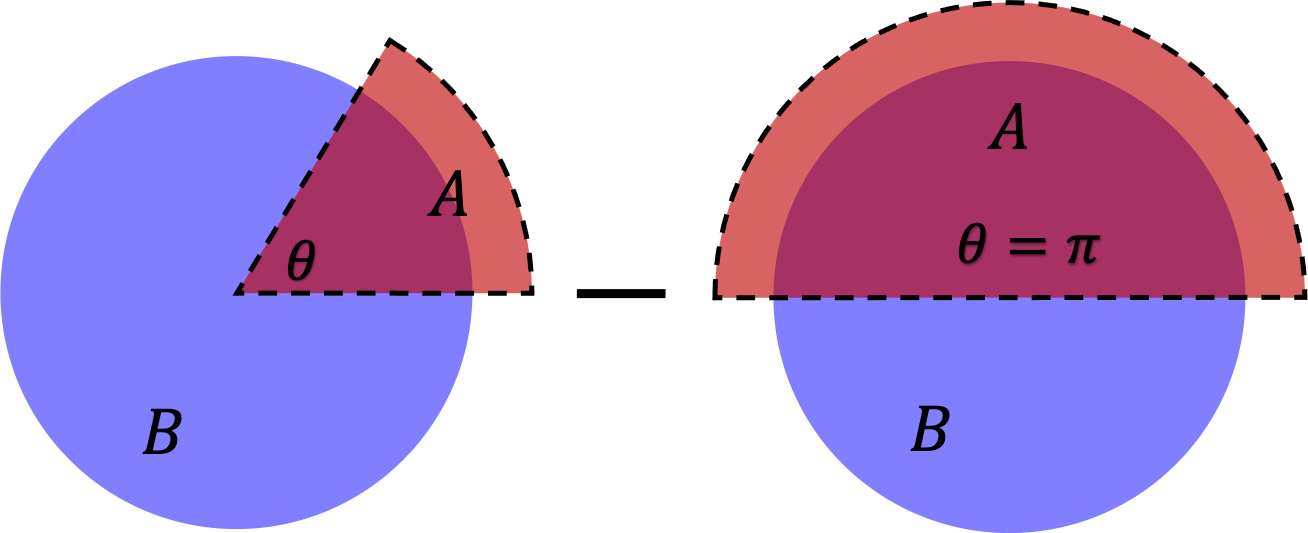}}
\caption{\label{fig8} The same cutting scheme as Fig.~\ref{fig2} with the subsystem A containing the physical boundary. Similar subtraction rule is applied to extract the corner and edge contribution.}
\end{figure}

If we extend the region A in the radial direction to infinity and thus its arc-shape boundary is the physical edge of the system. As shown in Fig.~\ref{fig8}, now the EE  between A and B contains the area law contribution from the radial boundary, a $\theta$-angle contribution at the center, two $\frac{\pi}{2}$-angle contributions at the intersections and the edge contribution on the boundary, namely
\begin{eqnarray}
S_\alpha = b_\alpha l_A + S_{\alpha}(\theta) + 2 S_{\alpha}(\pi/2) + S_{edge}  + \mathcal{O}(1/l_A).
\end{eqnarray}

\begin{figure}[ht]
\center{\includegraphics[width=7cm]  {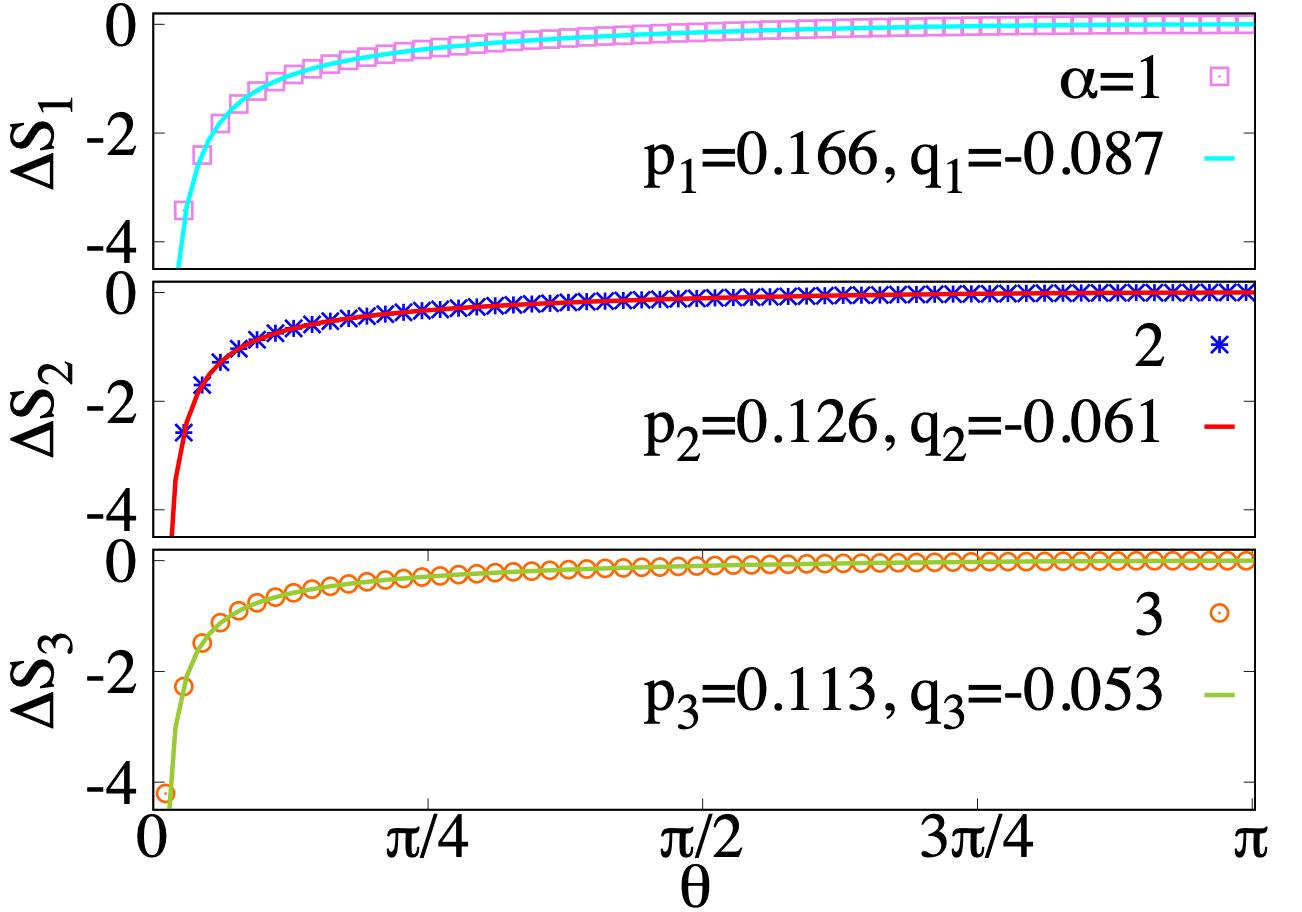}}
\caption{\label{fig9} The $\theta$-dependence of corner-edge contribution for R\'enyi entropy.  There data are fitted by $p_{\alpha}\ln \sin\frac{\theta}{2}+q_{\alpha}[1+(\pi-\theta)\cot\theta] $ and the results are $p_1=0.166(\pm 0.006)$, $q_1=-0.087(\pm 0.001)$, $p_2=0.126(\pm 0.001)$, $q_2=-0.061(\pm 0.0002)$, $p_3=0.113(\pm 0.0007)$, $q_3=-0.053(\pm 0.0001)$. }
\end{figure}

Similar to the previous section, after subtracting the EE at $\theta = \pi$, the area law term and $S_{\alpha}(\pi/2)$ are eliminated. Now the residual EE contains the corner and edge contributions. $\Delta S = S_{\alpha}(\theta) +  S_{edge}(\theta) - S_{edge}(\pi)$. The CFT predicts~\cite{Estienne20,Calabrese09,Berthiere19} that the EE of chiral edge is equal to $S_{edge}(\theta)  = \frac{c}{12} (1+\frac{1}{\alpha}) \ln [2R_A \sin(\theta/2)]$  where $2R_A \sin(\theta/2)$ is the chordal distance and the central charge $c = 1$. Therefore, we expect
\begin{eqnarray}
\Delta S_{\alpha} &=& S_{\alpha}(\theta) +  \frac{c}{12} (1+\frac{1}{\alpha}) \ln \sin\frac{\theta}{2}
\end{eqnarray}
which is independent of the $R_A$. Here we suppose the previous results of the corner keep invariant as  $S(\theta) = u_{\alpha}[1+(\pi-\theta)\cot\theta]$ and fit the data of $\Delta S$ by function $p_{\alpha}\ln \sin\frac{\theta}{2}+q_{\alpha}[1+(\pi-\theta)\cot\theta] $ with parameters $q_{\alpha}$ and $p_{\alpha}$. The results are shown in Fig.~\ref{fig9}.  The result $q_\alpha \simeq u_\alpha$ is expected which shows again that the corner and edge terms are independent.  The fitting results $p_1 = 0.166$, $p_2 = 0.126$ and $p_3 = 0.113$ are consistent with the formula $p_{\alpha} = \frac{1}{12} (1+\frac{1}{\alpha})$ and thus $c = 1$ is verified.

\begin{figure}[ht]
\center{\includegraphics[width=8cm]  {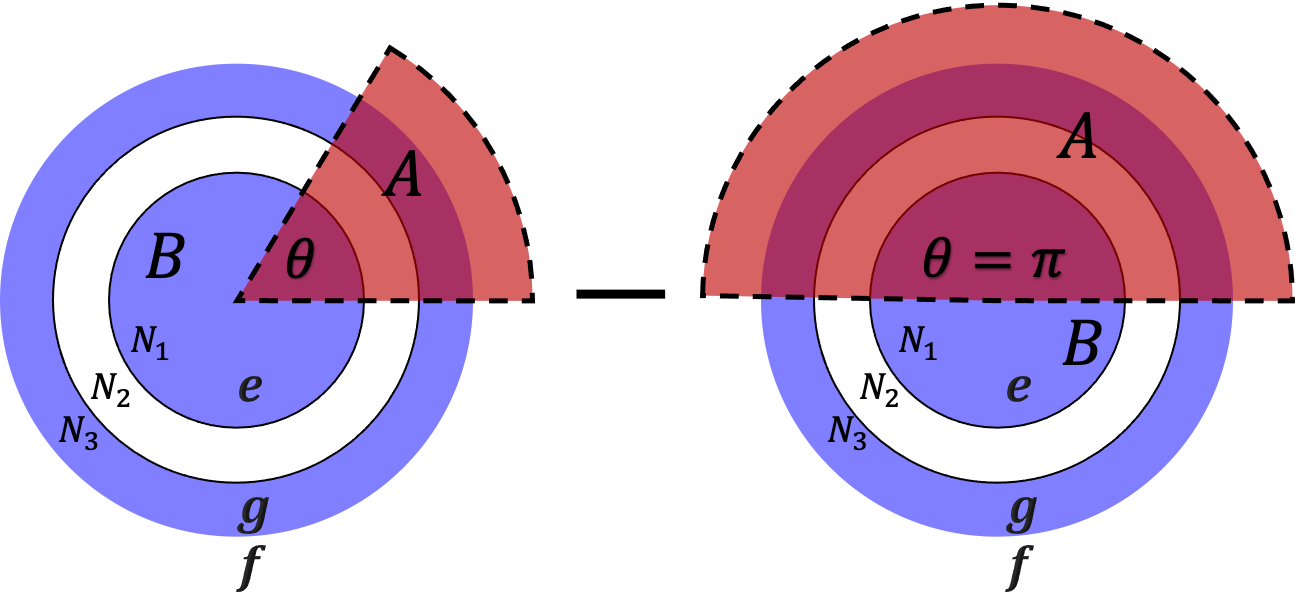}}
\caption{\label{fig10} The non-smooth cutting schematic diagram of a disk that occurs edge reconstruction. The system contains two unconnected part containing $N_1$ electrons in bulk and $N_3$ electrons in edge respectively, and they are separated by $N_2$ orbits. There are three chiral edge modes, and the inner edge of the reconstructed stripe has opposite chirality to that of the other two edges. By the same subtraction rule, we extract pure angle contribution and edge contribution.}
\end{figure}

To see how robust of the edge contribution of EE, we consider the edge reconstruction~\cite{Wen94, Wan02,Wan03} pattern as shown in Fig.~\ref{fig10}. The system contains two unconnected parts. One is the IQH state at the center with $N_1$ electrons, the other is the reconstructed part with $N_3$ electrons which we assume that they form the same IQH state at $\nu = 1$.  In the middle, there are $N_2$ unoccupied orbitals. In this case, although there are three chiral edge modes,  the inner edge of the reconstructed stripe has opposite chirality to that of the other two edges. Therefore, the total chirality is not affected by the edge reconstruction. We follow the same logic of subtracting the EE at $\theta = \pi$ as the unreconstructed case. We fix the total number of orbitals $N=N_1+N_2+N_3 = 171$ and consider several combinations of different $\{N_i\}$.  The fitting parameters are shown in Table.~\ref{table:1}. Here we only shows the results of $S_1$ and the results for other $S_{\alpha}$s could be expected.  It shows that the $p_1$ and $q_1$ are very robust and consistent to the previous unreconstructed results.

\begin{table}[ht]
\begin{center}
\caption{The corner and edge contribution for $S_1$ of a finite disk in edge reconstructed pattern. We fix the total number of orbitals $N=N_1+N_2+N_3 = 171$ and consider several combinations of different $\{N_i\}$. It shows that the $p_1$ and $q_1$ are very robust and consistent to the previous unreconstructed results.}
\label{table:1}
\begin{tabular}{|m{0.5cm}<{\centering}|m{0.5cm}<{\centering}|m{0.5cm}<{\centering}|m{3cm}<{\centering}|m{3cm}<{\centering}|}
\hline    \textbf{$N_1$} & \textbf{$N_2$} &\textbf{$N_3$} &\textbf{$p_1$} &\textbf{$q_1$} \\
\hline    151 & 10 & 10 & 0.16394$\pm 0.001172$ & -0.088024$\pm 0.000206$ \\
\hline    131 & 20 & 20 & 0.16810$\pm 0.001322$ & -0.087140$\pm 0.000232$ \\
\hline    141 & 10 & 20 & 0.16494$\pm 0.001169$ & -0.087827$\pm 0.000205$ \\
\hline    71 & 50 & 50 & 0.16835$\pm 0.001437$ & -0.087288$\pm 0.000290$ \\
\hline    41 & 50 & 80 & 0.16441$\pm 0.001914$ & -0.088775$\pm 0.000444$ \\
\hline
\end{tabular}
\end{center}
\end{table}

\section{Discussions and Conclusions}
As a conclusion, in the IQH state on a finite disk, we use a simple unified bipartite method to explore the independent EE contributions from the area law, the sharp corner and the gapless chiral edge contributions. The coefficients of the area law $b_{\alpha}$ is found to be universal and analytically solvable. With the exact area law term, we obtain the angle dependence of the corner contribution. It has a fixed prefactor and the behaviors at $\theta \rightarrow 0$ are consistent to the tip charge accumulation of a realistic IQH liquid on a cone surface. It is similar to a recent work of Ref.~\onlinecite{Estienne22} in which the $\theta$ dependence of particle number fluctuations at the corner was found to be universal and has the the same $1/\theta$ behavior while $\theta \rightarrow 0$.  While the fan-shaped subsystem containing the edge of disk, the gapless chiral quantum Hall edge contributes a logarithmic type of EE  in which the central charge $c$ in its prefactor is as expected by its underlying CFT.  Moreover, we find the edge reconstruction of the IQH dose not change any of the prefactors due to the conservation of the chirality.

Here we should note that the correlation matrix method only applicable to the non-interacting case, such as the IQH state. For the interacting case, namely the fractional quantum Hall (FQH) states, the direct calculation of the reduced density matrix with breaking the rotational symmetry on a disk is complicate and limited for small system size.  However, as was expected from the charge fluctuation calculations, we believe that our bipartite method is also applicable and the corner contribution in the FQH states still obey the same universality, especially in the limit $\theta \rightarrow 0$ where the charge densities at the tip are the same. The FQH edge also contributes a logarithmic type of EE which has a prefactor with its corresponding central charge.

\acknowledgments
This work was supported by National Natural Science Foundation of China Grant No. 11974064, the Chongqing Research Program of Basic Research and Frontier Technology Grant No. cstc2021jcyjmsxmX0081, Chongqing Talents: Exceptional Young Talents Project No. cstc2021ycjh-bgzxm0147, and the Fundamental Research Funds for the Central Universities Grant No. 2020CDJQY-Z003. QL is supported by National Natural Science Foundation of China Grant No. 61988102.

\end{document}